\documentclass[%
preprint,
amsmath,
amssymb,
aps,
prb,
]{revtex4-2}

\usepackage{graphicx}
\usepackage{dcolumn}
\usepackage{bm}
\usepackage[mathlines]{lineno}
\usepackage{comment}
\usepackage{color}
\usepackage{soul}
\sethlcolor{white}
\usepackage{ulem}
\usepackage{mathcomp}

\newcommand{\mr}[1]{\mathrm{#1}}

\begin{document}

\preprint{APS/123-QED}

\title{Modulated Dirac bands and integer hopping ratios in a honeycomb lattice of phenalenyl-tessellation molecules}
\author{Naoki Morishita}
\affiliation{%
 Research and Development Directorate, Japan Aerospace Exploration Agency, Tsukuba, Ibaraki 305-8505, Japan}%

\author{Kenshin Komatsu}
\author{Motoharu Kitatani}
\author{Koichi Kusakabe}
\affiliation{Graduate School of Science, University of Hyogo, Kamigori, Hyogo 678-1297, Japan}%

\date{\today}

\begin{abstract}
A family of nanographene molecules called phenalenyl-tessellation molecules (PTMs) exhibits two types of zero modes: 
a $\sqrt{3} \times \sqrt{3}$ type that spreads over the entire molecule and a vacancy-localized type. 
A periodic system of PTMs is expected to have low-energy bands that strongly reflect the properties of the zero modes of PTMs as effective atoms.
In this study, we show that the low-energy Dirac bands in a class of honeycomb PTMs (H--PTM) can be represented by an effective honeycomb model 
which is determined only by the connections between neighboring effective atoms.
The hopping parameters of H--PTM in each direction 
take positive integer ratios according to the connection order between two PTMs.
By structurally designing each PTM, we can change the connection order of the PTMs and hence modulate the energy gap and the Fermi velocity of the Dirac band of the H--PTM.
Moreover, we confirm that Dirac bands coexist with vacancy-localized zero modes in the H--PTM with vacancies.
The result indicates that the nanographene structure arranging PTMs as effective atoms extends material design freedom that effectively generates a modulated Dirac electron system with coexisting localized electron spins for graphene-based electronic and quantum devices.
\end{abstract}

\maketitle

\section{\label{Intro}Introduction} 
Graphene, already applied to electronic devices\cite{Novoselov666, Novoselov2005, Geim2007, Novoselov2012, Wolf2014, MBAYACHI2021100163, Kireev2021}, 
holds the potential for applications in quantum devices due to its zero modes and localized spins at specific edges\cite{doi:10.1143/JPSJ.65.1920, PhysRevB.54.17954, PhysRevB.84.115406, Ziatdinov2017, doi:10.7566/JPSJ.87.084706, doi:10.1021/acsnano.0c01737, Wang2021} and vacancies\cite{PhysRevLett.96.036801, PhysRevB.77.115109, PhysRevLett.93.187202, PhysRevB.89.155405, doi:10.7566/JPSJ.85.084703, doi:10.7566/JPSJ.88.124707, MORISHITA2021PLA, Morishita_APEX_2021}. 
There are also theoretical proposals for nanostructures that give various low-energy effective models,
such as Dirac bands and localized flat bands, for future applications that take advantage of their unique transport properties.\cite{PhysRevB.101.205311, doi:10.1021/acs.nanolett.8b04616}
To discuss low-energy modes, two-dimensional structures are often proposed from open-shell molecules.\cite{PhysRevB.108.115113, Ortiz_2023, PhysRevResearch.6.043262} 
Therefore, a general method to design a co-existent system of Dirac bands and localized zero modes at will in graphene nanostructures based on molecular structures 
would significantly widen the applicability range of graphene.
 
As is well known, 
the isotropic Dirac cone in a two-site nearest-neighbor tight-binding (TB) model 
can be modulated if the value of the hopping parameters between the A and B sites 
were varied depending on the direction of the three $\pi$-bonds\cite{Hasegawa_PRB_2006, PhysRevB.93.035439}.
Specifically, 
when the values of the hopping parameters $t_{0}, t_{1}$ and $t_{2}$ 
in the three directions satisfy the triangle inequality, 
there is no band gap; 
however, if the triangle inequality is violated, a band gap is formed\cite{Hasegawa_PRB_2006, Kishigi_2011}.
Although anisotropic hopping parameters can be induced in graphene, 
for example, by applying mechanical strain\cite{PhysRevLett.103.046801, PAPAGEORGIOU201775}, 
deformations beyond 20\% are required for band-gap opening\cite{PhysRevB.80.045401, doi:10.1021/nn800459e, doi:10.1021/nn8008323},
which implies precisely controlling the pairs of $t_{0}, t_{1}$ and $t_{2}$ to any arbitrary ratio is difficult.

Recently, we have been closely studying zero modes appearing in phenalenyl-tessellation molecules (PTMs), 
in which the molecular structure can be defined by tiling phenalene’s carbon skeleton as a unit structure.\cite{doi:10.7566/JPSJ.88.124707,MORISHITA2021PLA}
PTMs possess a unique property that provides significant advantages over the other design approaches.
In molecules composed of multiple PTMs, the number of $\sqrt{3} \times \sqrt{3}$ type zero mode appearances are governed by the number of connections between PTMs.
In addition, vacancy-localized zero modes can be introduced by placing vacancies at appropriate positions that match the shape of the $\sqrt{3} \times \sqrt{3}$ type zero mode, 
enabling the design of high-spin systems, such as $S=1$ and $S=3/2$ with multiple electron spins.\cite{Morishita_APEX_2021, KomatsuJPSJ2025}.

In this context, we propose a general method for constructing nanographene structures based on the PTM structure, 
in which the effective hopping parameters of the low-energy band in a two-site honeycomb system can be controlled in a positive integer ratio by the number of connections between PTMs.
In addition to enabling control over the effective hopping parameters of the low-energy bands and the band gap of the system, 
the proposed method also allows vacancy-derived localized zero modes to coexist with modulated Dirac electrons.
The effectiveness of the proposed method for modulation of the Dirac bands is numerically confirmed in structural examples. 
The controllability of the Fermi velocity and the coexistence of Dirac dispersion and localized zero modes are also discussed.

\section{\label{Method}Method and theory}
\subsection{\label{Method_Definition}Definition of the honeycomb model}

\begin{figure}
\includegraphics[width=0.5\linewidth]{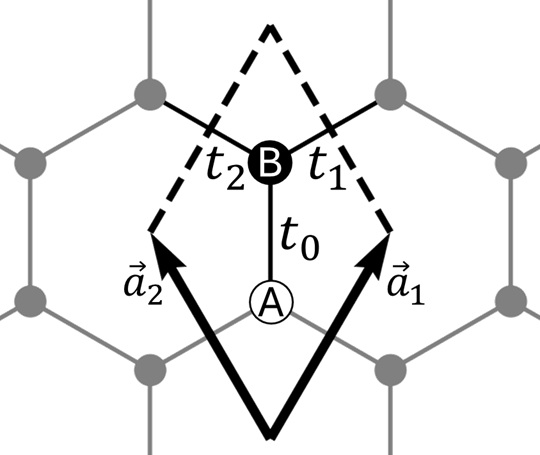}
\caption{\label{Fig1} 
A honeycomb lattice with an A-site and a B-site. $t_0$, $t_1$ and $t_2$ are the hopping parameters in the three directions.}
\end{figure}
We first review the graphene structure of a honeycomb lattice with two sites, 
named the A-site and the B-site, as shown in Fig.~\ref{Fig1}.
There are $\pi$ bonds in three directions between the A-site and the B-site, 
and the hopping parameter of each bond can be expressed as $t_0$, $t_1$ and $t_2$.
In pristine graphene, $t_0 = t_1 = t_2 = t$ ($\sim$2.7 eV\cite{RevModPhys.81.109}), 
and the length of the lattice vectors $\vec{a}_{1} = (a_{1x}, a_{1y})$ and $\vec{a}_{2} = (a_{2x}, a_{2y})$ are $a$ ($\sim$2.46 \AA).

\begin{figure}
\includegraphics[width=0.6\linewidth]{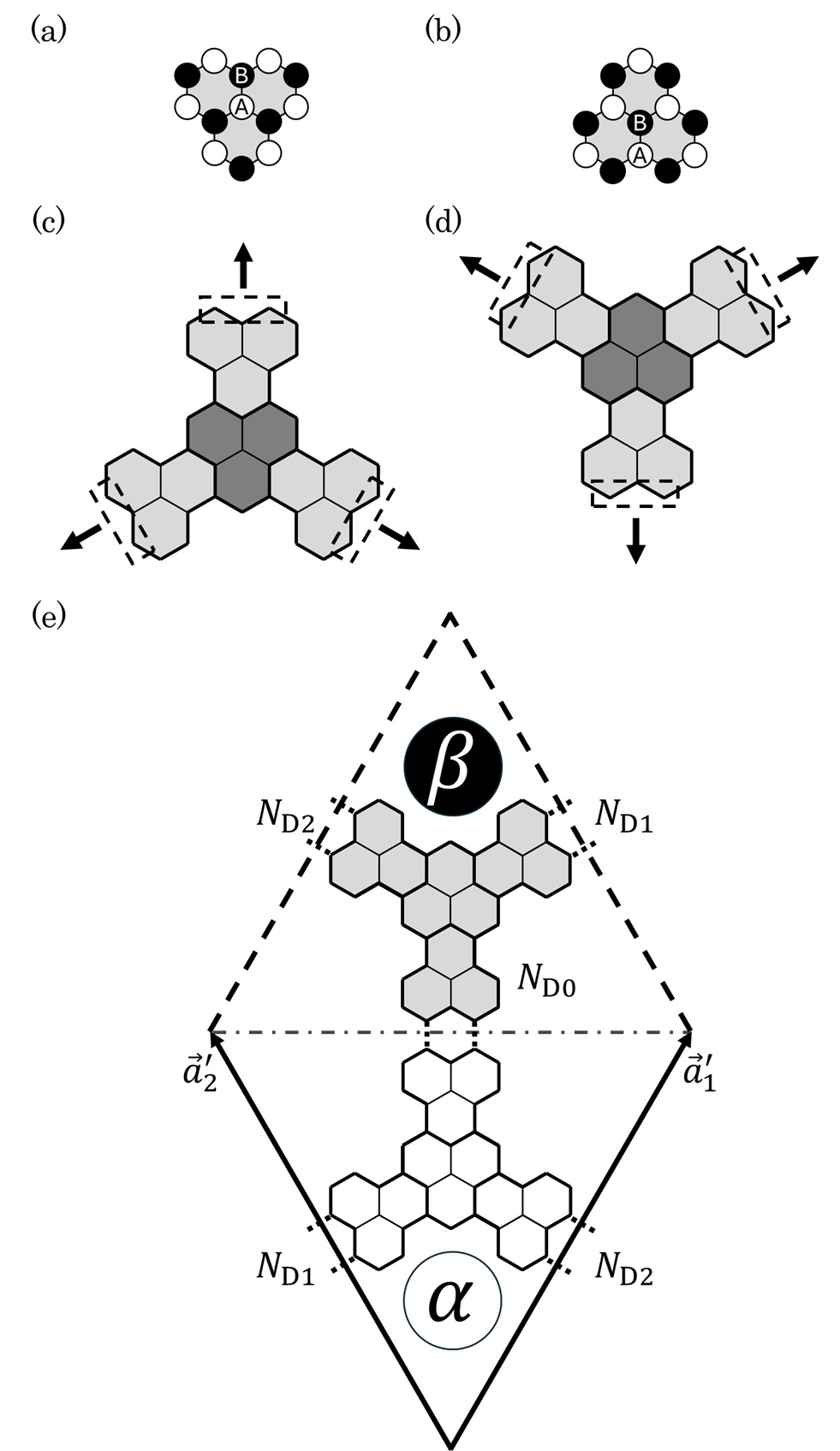}
\caption{\label{Fig2} (a) An $\alpha$-phenalenyl unit (PU), 
(b) a $\beta$-PU, 
(c) an $\alpha$-phenalenyl-tessellation molecule (PTM) with four $\alpha$-PUs, and 
(d) a $\beta$-PTM with four $\beta$-PUs. 
Light gray and dark gray denote the PUs. 
Dashed rectangles indicate the double-zigzag corners (DZCs) of the PTMs.  
(e) A honeycomb PTMs (H--PTM) structure with $N_{\mr{D}0}=N_{\mr{D}1}=N_{\mr{D}2}=1$.
The $\alpha$-PTM (lower half) and the $\beta$-PTM (upper half) are connected through the dashed lines at DZCs.
In the subsequent figures, the border between the $\alpha$-PTM and the $\beta$-PTM within a unit cell is indicated by a horizontal dashed-dotted line.}
\end{figure}

We now define a nanographene effective atom structure. 
An effective atom is given by a PTM\cite{doi:10.7566/JPSJ.88.124707, MORISHITA2021PLA} which is a specifically defined nanographene molecule.
A PTM is composed of a tessellated skeleton of phenalene molecules. 
Here, a phenalenyl unit (PU) is defined as a bipartite graph of seven A-sites and six B-sites (or six A-sites and seven B-sites).
Figures ~\ref{Fig2}(a) and ~\ref{Fig2}(b) show PUs with A site and B site as the central site, respectively.
Hereafter the former PU is defined as $\alpha$-PU and the latter as $\beta$-PU, respectively.
A PTM comprises one or more PUs. 

We further consider a nanographene honeycomb structure consisting of two PTMs in a unit cell.
We denote the number of PUs  that constitute a PTM as $N_{\mr{PU}}$. 
As examples, we consider the PTMs composed of four PUs shown in Fig.~\ref{Fig2} (c) and (d). 
In these molecules, $N_{\mr{PU}} =4$. 
The central site of each PU in Fig.~\ref{Fig2} (c) and (d) is the A-site and B-site shown in Fig.~\ref{Fig2} (a) and (b), respectively. 
Hereafter, we refer to a PTM with $\alpha$-PUs and $\beta$-PUs  as an $\alpha$-PTM and $\beta$-PTM, respectively.
The corners of PTMs possess two consecutive zigzag structures, referred to as double-zigzag corners (DZCs).
The $\alpha$-PTM and $\beta$-PTM structures shown in Fig.~\ref{Fig2} (c) and (d), respectively, have three DZCs in three directions.  
We thus define a periodic honeycomb structure 
that connects an $\alpha$-PTM and a $\beta$-PTM via DZCs in three directions 
with lattice vectors $\vec{a}'_{1} = (a'_{1x}, a'_{1y})$ and $\vec{a}'_{2} = (a'_{2x}, a'_{2y})$ (see Fig.~\ref{Fig2} (e)). 
We call this structure a \textit{honeycomb PTMs (H--PTM)}.
Let us denote the number of DZCs connected in each direction as $N_{\mr{D}0}$, $N_{\mr{D}1}$ and $N_{\mr{D}2}$. 
In the present example, $N_{\mr{D}0}=N_{\mr{D}1}=N_{\mr{D}2}=1$.

\subsection{\label{Method_Hamiltonian}Hamiltonian of an H--PTM}
Let us consider the tight-binding (TB) Hamiltonian of a two-site honeycomb lattice with 
nearest-neighbor hopping parameters 
$t_{0}=T_{0}t$, $t_{1}=T_{1}t$ and $t_{2}=T_{2}t$ 
along the three directions.
The TB Hamiltonian can be written in $2 \times 2$ matrix form as follows:
\begin{equation}
\begin{split}
\label{twosite}
&H_{\mr{two}}(\vec{k}) \\
&=
t 
\begin{pmatrix}
     0 & T_{0} + T_{1}e^{-i \vec{k} \cdot \vec{a}_{1}} + T_{2}e^{-i \vec{k} \cdot \vec{a}_{2}}\\
     T_{0} + T_{1}e^{i \vec{k} \cdot \vec{a}_{1}} + T_{2}e^{i \vec{k} \cdot \vec{a}_{2}} & 0
\end{pmatrix}.
\end{split}
\end{equation}
This expression accounts for the phase factor, which depends on the wavenumber $\vec{k}$,
when hopping over the boundary of the unit cell.

Next, we consider the Hamiltonian of an H--PTM 
in which unit cell is composed of one $\alpha$-PTM and one $\beta$-PTM.
In this section, we assume that both an $\alpha$-PTM and a $\beta$-PTM themselves contain no internal vacancies.
Note that, although cavities surrounded by 
periodically arranged $\alpha$-PTMs and $\beta$-PTMs are formed, 
the shape and size of these external cavities do not affect the following discussion.

The number of A-sites and B-sites in the $\alpha$-PTM region (the part below the dashed line that bisects the unit cell in Fig.~\ref{Fig2} (e)), 
denoted by $N_{\mr{A}\alpha}$ and $N_{\mr{B}\alpha}$, respectively, satisfies $N_{\mr{A}\alpha} - N_{\mr{B}\alpha} = 1$.
A similar relationship, $N_{\mr{B}\beta} - N_{\mr{A}\beta} = 1$, holds in the $\beta$-PTM region (the part above the dashed line in Fig.~\ref{Fig2} (e)).
The total number of A-sites and B-sites in a unit cell is 
$N_{\mr{A}} = N_{\mr{A}\alpha} + N_{\mr{A}\beta}$ and $N_{\mr{B}} = N_{\mr{B}\alpha} + N_{\mr{B}\beta}$, respectively.
Therefore, $N_{\mr{A}} = N_{\mr{B}}$ holds 
throughout the unit cell.
The total number of sites in the unit cell is $N = N_{\mr{A}}+N_{\mr{B}}$. 
Accordingly, the TB Hamiltonian of the H--PTM can be written in $N \times N$ matrix form as  
\begin{equation}
\label{Ham_H--PTM}
H_{\mr{H-PTM}}(\vec{k}) =
  \begin{pmatrix}
     0 & H_{\mr{AB}}(\vec{k}) \\
     H_{\mr{BA}}(\vec{k}) & 0
  \end{pmatrix}.
\end{equation}
This Hamiltonian is composed of a real part describing hopping within a unit cell 
and a phase factor considering the $k$-dependent hopping across the unit cell.
Namely, 
$H_{\mr{AB}}(\vec{k}) = H_{\mr{AB}}^{0} + e^{-i \vec{k} \cdot \vec{a}'_{1}}H_{\mr{AB}}^{a_{1}} + e^{-i \vec{k} \cdot \vec{a}'_{2}}H_{\mr{AB}}^{a_{2}}$, 
where $H^{0}_{\mr{AB}}$ is the real block matrix having $t$ in its $i, j$-th elements 
when sites $i$ and $j$ are adjacent within the unit cell, 
and 
$H_{\mr{AB}}^{a_{1}}$ and $H_{\mr{AB}}^{a_{2}}$ are real block matrices
having $t$ in their $i, j$-th elements when sites $i$ and $j$ are adjacent
across the unit cell in the $\vec{a}'_{1}$ and $\vec{a}'_{2}$-directions, respectively.
$H_{\mr{BA}}(\vec{k})$ is the conjugate transpose of $H_{\mr{AB}}(\vec{k})$.

\subsection{\label{Method_BasisChange}Hamiltonian transformation} 
Because the system is bipartite with $N$ sites and the eigenvalues (energy bands) of the TB Hamiltonian of an H--PTM are symmetric with respect to the Fermi level ($E=0$), 
we can arrange the eigenvalue pairs $\pm E_{n}(\vec{k}) (n = 1, ..., N/2)$ in ascending order of their absolute values and corresponding eigenvectors $\vec{\Psi}_{\pm n}(\vec{k})$.
The orthogonal eigenvectors $\vec{\Psi}_{\pm n}(\vec{k})$ are expressed as follows:
\begin{equation}
 \vec{\Psi}_{\pm n}(\vec{k})  = (\vec{\psi}_{\mr{A}n}(\vec{k})  \pm \vec{\psi}_{\mr{B}n}(\vec{k}) )/\sqrt{2},
\end{equation}
where $\vec{\psi}_{\mr{A}n}(\vec{k})$ and $\vec{\psi}_{\mr{B}n}(\vec{k})$ are the basis vectors on the A-sites and B-sites, respectively.
With these basis vectors in ascending order, 
one can write down the unitarily transformed Hamiltonian as follows:
\begin{equation}
H'_{\mr{H-PTM}}(\vec{k})  = 
  \begin{pmatrix}
    0 & H'_{\mr{AB}}(\vec{k}) \\
    H'_{\mr{BA}}(\vec{k})  & 0
  \end{pmatrix},
\end{equation}
where each element of the block matrix $H'_{\mr{AB}}(\vec{k})  = [t'_{nm}(\vec{k}) ]$ is given by 
\begin{equation}
\label{Eq_tnn}
  t'_{nm}(\vec{k})  = 
  \begin{cases}
\vec{\psi}^{*}_{\mr{A}n}(\vec{k})  H_{\mr{AB}}(\vec{k})  \vec{\psi}_{\mr{B}n}(\vec{k}) & (m=n), \\
 0 & (m \neq n).
\end{cases} 
\end{equation}
and $H'_{\mr{BA}}(\vec{k}) $ is the conjugate transpose of $H'_{\mr{AB}}(\vec{k}) $.
Note that as the eigenvectors are orthogonal, 
the off-diagonal elements of $H'_{\mr{AB}}(\vec{k}) $ are all zero. 

\subsection{\label{Method_Effective_t}Low-energy effective model}
In this section, we analyze the low-energy bands.
To obtain the eigenvalues $E_{\pm n}(\vec{k})$ of $H'_{\mr{H-PTM}}(\vec{k})$, we need only to find the eigenvalues of the following $2 \times 2$ Hamiltonian:
\begin{equation}
\label{Hpmn}
H_{\pm n}(\vec{k}) = \begin{pmatrix}
     0 & t'_{nn}(\vec{k}) \\
     {t'^{*}_{nn}}(\vec{k}) & 0
  \end{pmatrix}.
\end{equation}
Focusing on $E_{\pm 1}(\vec{k})$, the energy eigenvalue closest to the Fermi level of 
the $2 \times 2$ Hamiltonian obtained from Eqs.~(\ref{Eq_tnn}) and (\ref{Hpmn}) becomes
\begin{equation}
\label{Hpm1}
\begin{split}
&H_{\pm 1}(\vec{k}) \\
&=
    \begin{pmatrix}
     0 & \vec{\psi}^{*}_{\mr{A}1}(\vec{k}) H_{\mr{AB}}(\vec{k}) \vec{\psi}_{\mr{B}1}(\vec{k}) \\
(\vec{\psi}^{*}_{\mr{A}1}(\vec{k}) H_{\mr{AB}}(\vec{k}) \vec{\psi}_{\mr{B}1}(\vec{k}))^{*} & 0
  \end{pmatrix},
\end{split}
\end{equation}
where $\vec{\psi}_{\mr{A}1}(\vec{k})$ and $\vec{\psi}_{\mr{B}1}(\vec{k})$ can be divided into components within the $\alpha$-PTM and $\beta$-PTM regions as $\vec{\psi}_{\mr{A}1}(\vec{k})=\vec{\psi}_{\mr{A}1 \alpha}(\vec{k})+\vec{\psi}_{\mr{A}1 \beta}(\vec{k})$ and $\vec{\psi}_{\mr{B}1}(\vec{k})=\vec{\psi}_{\mr{B}1 \alpha}(\vec{k})+\vec{\psi}_{\mr{B}1 \beta}(\vec{k})$, respectively.
We then have, 
$\vec{\psi}_{\mr{A}1}^{*}(\vec{k}) H_{\mr{AB}}(\vec{k}) \vec{\psi}_{\mr{B}1}(\vec{k}) 
= [\vec{\psi}_{\mr{A}1 \alpha}^{*}(\vec{k})+\vec{\psi}_{\mr{A}1 \beta }^{*}(\vec{k})] 
H_{\mr{AB}}(\vec{k})
[\vec{\psi}_{\mr{B}1 \alpha}(\vec{k}) +\vec{\psi}_{\mr{B}1 \beta}(\vec{k})] \\
=
\vec{\psi}_{\mr{A}1 \alpha}^{*}(\vec{k}) H_{\mr{AB}}(\vec{k}) \vec{\psi}_{\mr{B}1 \beta}(\vec{k})
+
\vec{\psi}_{\mr{A}1 \alpha}^{*}(\vec{k}) H_{\mr{AB}}(\vec{k}) \vec{\psi}_{\mr{B}1 \alpha}(\vec{k})
+
\vec{\psi}_{\mr{A}1 \beta}^{*}(\vec{k}) H_{\mr{AB}}(\vec{k}) \vec{\psi}_{\mr{B}1 \beta}(\vec{k})
+
\vec{\psi}_{\mr{A}1 \beta}^{*}(\vec{k}) H_{\mr{AB}}(\vec{k}) \vec{\psi}_{\mr{B}1 \alpha}(\vec{k})$.
The last term $\vec{\psi}_{\mr{A}1 \beta}^{*}(\vec{k}) H_{\mr{AB}}(\vec{k}) \vec{\psi}_{\mr{B}1 \alpha}(\vec{k})$ vanishes
because $\vec{\psi}_{\mr{A}1 \beta}(\vec{k})$ and $\vec{\psi}_{\mr{B}1 \alpha}(\vec{k})$ contain no adjacent sites. 

\begin{figure}
\includegraphics[width=1.0\linewidth]{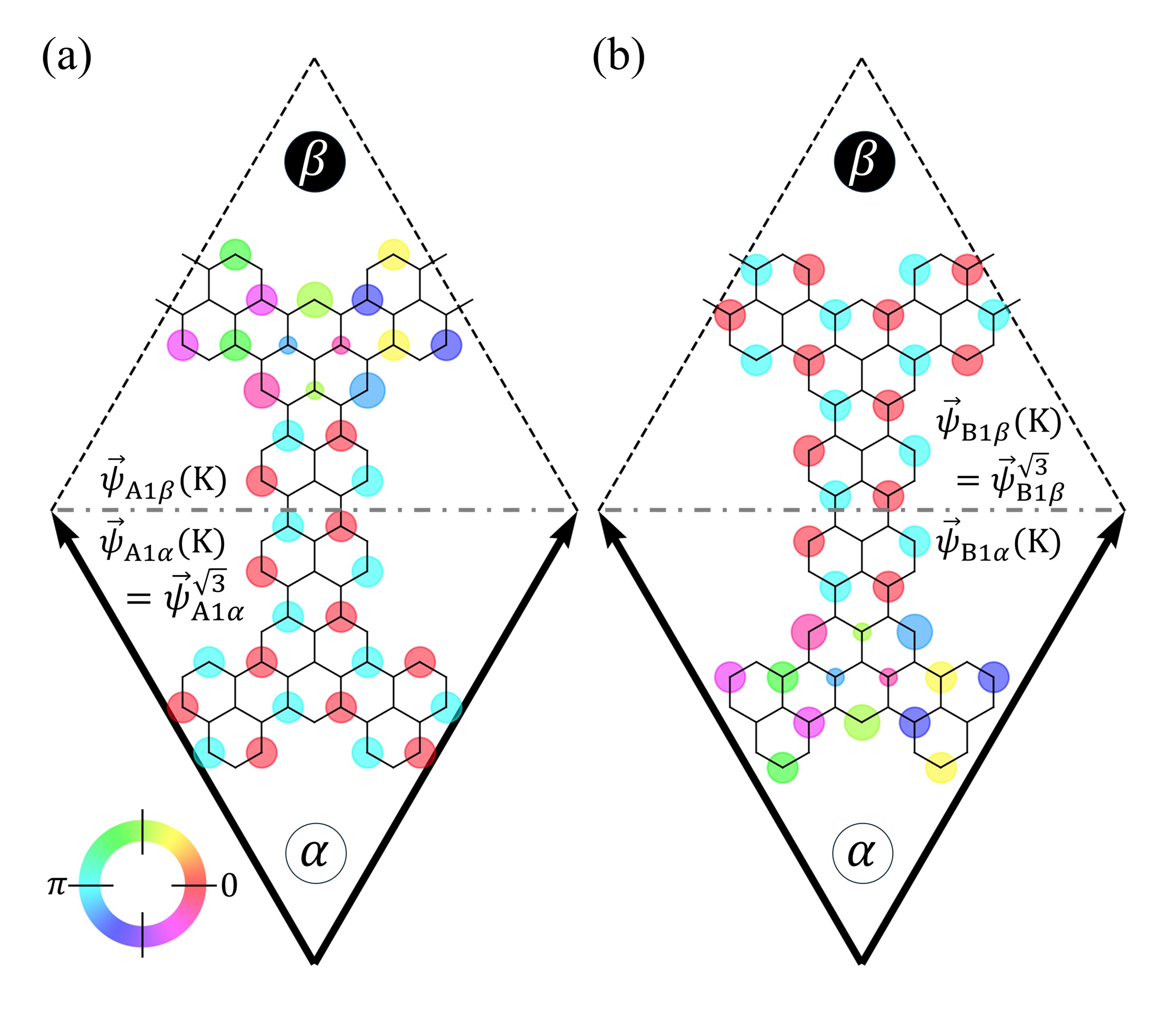}
\caption{\label{Fig3} Eigenvectors at the K-point on (a) the A-sites 
$\vec{\psi}_{\mr{A}1}(\mr{K})=\vec{\psi}_{\mr{A}1 \alpha}(\mr{K})+\vec{\psi}_{\mr{A}1 \beta}(\mr{K})=\vec{\psi}_{\mr{A}1 \alpha}^{\sqrt{3}}+\vec{\psi}_{\mr{A}1 \beta}(\mr{K})$ and (b) B-sites
$\vec{\psi}_{\mr{B}1}(\mr{K})=\vec{\psi}_{\mr{B}1 \alpha}(\mr{K})+\vec{\psi}_{\mr{B}1 \beta}(\mr{K})=\vec{\psi}_{\mr{B}1 \alpha}(\mr{K})+\vec{\psi}_{\mr{B}1 \beta}^{\sqrt{3}}$. 
The hue (lower left) represents the phase.}
\end{figure}

We now consider an approximation 
in the vicinity of $E=0$. 
By assuming $E=0$ and applying the zero-sum rule for zero modes \cite{Borden_1977, Cash_2001, Langler_2002, Orimoto_2006} to $\vec{\psi}_{\mr{A}1 \alpha}(\vec{k})$ and $\vec{\psi}_{\mr{B}1 \beta}(\vec{k})$, a
$\sqrt{3} \times \sqrt{3}$ shape is imposed on the $\alpha$-PTM and $\beta$-PTM regions 
as seen in Fig.~\ref{Fig3} (a) and (b), respectively.
Under the zero sum rule, the wavefunction values of the B(A) sites adjacent to an A(B) site sum to zero, as reported in molecular systems\cite{MORISHITA2021PLA}.
Expressing the $\sqrt{3} \times \sqrt{3}$-shaped 
$\vec{\psi}_{\mr{A}1 \alpha}$ and $\vec{\psi}_{\mr{B}1 \beta}$
as $\vec{\psi}_{\mr{A}1\alpha}^{\sqrt{3}}$ and $\vec{\psi}_{\mr{B}1\beta}^{\sqrt{3}}$, respectively, 
we have 
$\vec{\psi}_{\mr{A}1}^{*}(\vec{k}) H_{\mr{AB}}(\vec{k}) \vec{\psi}_{\mr{B}1}(\vec{k}) 
=
\vec{\psi}_{\mr{A}1\alpha}^{\sqrt{3}*} H_{\mr{AB}}(\vec{k}) \vec{\psi}_{\mr{B}1\alpha}^{\sqrt{3}}
+
\vec{\psi}_{\mr{A}1\alpha}^{\sqrt{3}*} H_{\mr{AB}}(\vec{k}) \vec{\psi}_{\mr{B}1 \alpha}(\vec{k})
+
\vec{\psi}_{\mr{A}1 \beta}^{*}(\vec{k}) H_{\mr{AB}}(\vec{k}) \vec{\psi}_{\mr{B}1\alpha}^{\sqrt{3}}
$.
In the example of an isotropic structure shown in Fig.~\ref{Fig2} (e), $E=0$ when $\vec{k}=$K ( or K').
The eigenvectors $\vec{\psi}_{\mr{A}1}(\mr{K})$ and $\vec{\psi}_{\mr{B}1}(\mr{K})$ at the A and B sites are shown in Fig.~\ref{Fig3} (a) and (b), respectively.

For uniformly distributed $\sqrt{3} \times \sqrt{3}$ modes 
$\vec{\psi}_{\mr{A}1 \alpha}^{\sqrt{3}}$ and $\vec{\psi}_{\mr{B}1 \beta}^{\sqrt{3}}$ with alternating signs, the second and third terms become 
$
\vec{\psi}_{\mr{A}1 \alpha}^{\sqrt{3}*} H_{\mr{AB}}(\vec{k}) \vec{\psi}_{\mr{B}1 \alpha}(\vec{k}) = 0$
and
$
\vec{\psi}_{\mr{A}1 \beta}^{*}(\vec{k}) H_{\mr{AB}}(\vec{k})\vec{\psi}_{\mr{B}1 \beta}^{\sqrt{3}} = 0$, respectively.
The first term can also be simplified for $\sqrt{3} \times \sqrt{3}$ modes.
As the $\sqrt{3} \times \sqrt{3}$ modes have uniform amplitudes, 
and the components of the vectors $\vec{\psi}_{\mr{A}1 \alpha}^{\sqrt{3}}$ and $\vec{\psi}_{\mr{B}1 \beta}^{\sqrt{3}}$ at each site can be written as real values with alternating signs; that is, as $\pm|{\varphi}_{\alpha}|$ and $\pm|{\varphi}_{\beta}|$, respectively.
Furthermore, as 
$\vec{\psi}_{\mr{A}1 \alpha}^{\sqrt{3}}$ and 
$\vec{\psi}_{\mr{B}1 \beta}^{\sqrt{3}}$ in adjacent
$\alpha$-PTMs and $\beta$-PTMs are connected with two bonds between each of their DZCs;  
their contribution to the total sum is only 
$2 {N_{\mr{D}0}} |{{\varphi}_{\alpha}}||{{\varphi}_{\beta}}| t 
+ 2 {N_{\mr{D}1}} |{{\varphi}_{\alpha}}||{{\varphi}_{\beta}}| t e^{-i \vec{k} \cdot \vec{a}'_{1}}
+ 2 {N_{\mr{D}2}} |{{\varphi}_{\alpha}}||{{\varphi}_{\beta}}| t e^{-i \vec{k} \cdot \vec{a}'_{2}}
$.

Therefore, the lowest-order estimation of the low-energy bands in $\sqrt{3} \times \sqrt{3}$ modes is
\begin{widetext}
\begin{equation}
\label{Heff}
H_{\mr{eff}} 
= \tau 
\begin{pmatrix}
     0 &  {N_{\mr{D}0}}
     +{N_{\mr{D}1}}e^{-i \vec{k} \cdot \vec{a}'_{1}} 
     +{N_{\mr{D}2}}e^{-i \vec{k} \cdot \vec{a}'_{2}} \\
      {N_{\mr{D}0}}
     +{N_{\mr{D}1}}e^{i \vec{k} \cdot \vec{a}'_{1}} 
     +{N_{\mr{D}2}}e^{i \vec{k} \cdot \vec{a}'_{2}}  & 0
\end{pmatrix},
\end{equation}
\end{widetext}
where 
\begin{equation}
\label{tau}
\tau =  2 |{{\varphi}_{\alpha}}||{{\varphi}_{\beta}}| t.
\end{equation}
Note that the value of ${\varphi}_{\alpha}$ is determined by the normalization factor of the eigenvectors satisfying $E=0$ in each structure. 
This value can be easily analytically determined for small or structurally simple systems (See Appendix for more details).

Equation~(\ref{Heff}) is an effective honeycomb model analogous to Eq.~(\ref{twosite})
with replacements $t \rightarrow \tau$, $T_{0} \rightarrow N_{\mr{D}0}$, $T_{1} \rightarrow N_{\mr{D}1}$, $T_{2} \rightarrow N_{\mr{D}2}$, $\vec{a}_{1} \rightarrow \vec{a}'_{1}$ and $\vec{a}_{2} \rightarrow \vec{a}'_{2}$.
Denoting the effective hopping parameters in the three directions as 
$\tau_{0} = N_{\mr{D}0}\tau$, $\tau_{1} = N_{\mr{D}1}\tau$ and $\tau_{2} = N_{\mr{D}2}\tau$, we find that
$\tau_{0}$, $\tau_{1}$ and $\tau_{2}$ have positive integer ratios.
Because the effective two-site system corresponding to Eq.~(\ref{Heff}) is bipartite, 
the two eigenvalues are symmetric with respect to the Fermi level and take values of $\pm E_{\mr{eff}}(\vec{k})$.

\subsection{\label{Method_GapFermiV}Band gap and Fermi velocity}
For general $\vec{k}$, we write
$\vec{\psi}_{\mr{A}1 \alpha}(\vec{k})$ and $\vec{\psi}_{\mr{B}1 \beta}(\vec{k})$ as 
$\vec{\psi}_{\mr{A}1 \alpha}(\vec{k}) = \vec{\psi}_{\mr{A}1 \alpha}^{\sqrt{3}} + \vec{\psi}'_{\mr{A}1 \alpha}(\vec{k})$
and 
$\vec{\psi}_{\mr{B}1 \beta}(\vec{k}) = \vec{\psi}_{\mr{B}1 \beta}^{\sqrt{3}} + \vec{\psi}'_{\mr{B}1 \beta}(\vec{k})$, respectively. 
Therefore, the original and effective band energies derived by Eqs.~(\ref{Hpm1}) and ~(\ref{Heff}), respectively,
differ in their higher-order terms.
However, when 
$\vec{\psi}'_{\mr{A}1 \alpha}(\vec{k}) = 0$ 
and 
$\vec{\psi}'_{\mr{B}1 \beta}(\vec{k}) = 0$, the effective Hamiltonian given in Eq.~(\ref{Heff}) yields an exact match  with Eq.~(\ref{Hpm1}) for $E=0$, 
providing accurate information on gap closure, 
the location of the $k$-point where the gap closes, 
and the differentiation of the band around the gap-closing point; that is, the Fermi velocity.
Conversely, if no eigenvalue of Eq.~(\ref{Heff}) satisfies $E=0$, 
then Eq.~(\ref{Hpm1}) has no eigenvalue $E=0$  
under the condition 
$\vec{\psi}_{\mr{A}1 \alpha}(\vec{k}) = \vec{\psi}_{\mr{A}1 \alpha}^{\sqrt{3}}$
and 
$\vec{\psi}_{\mr{B}1 \beta}(\vec{k}) = \vec{\psi}_{\mr{B}1 \beta}^{\sqrt{3}}$, 
the only case satisfying the zero-sum rule. 
Therefore, the existence of gaps in Eq.~(\ref{Heff}) corresponds to that in Eq.~(\ref{Hpm1}).

Let us briefly examine the behavior of the low-energy bands represented by the effective Hamiltonian.
In a honeycomb model with general hopping parameters 
$t_0$, $t_1$ and $t_2$,
the triangular inequality determines the existence or non-existence of a band gap. \cite{Hasegawa_PRB_2006, Kishigi_2011} 
If $t_0$, $t_1$ and $t_2$ 
satisfy the triangular inequality, the gap closes. 
Graphene is a well-known example in which $t_0 = t_1 = t_2 = t$ satisfies the triangle inequality. 
As is well-known, Dirac cones are formed in graphene, and the gaps are closed at the K and K' points.
Differentiating the isotropic Dirac cones, the Fermi velocity $\nu$ is given by:
\begin{equation}
  \label{EqDis1}
  \nu = 
  \frac{\sqrt{3}}{2} \frac{a t} {\hbar}.
\end{equation}
At the critical values for which the triangular inequality holds, 
e.g., $t_0 = t_1 + t_2$, 
a special band structure in which the tips merge with the anisotropic Fermi velocity at the confluent point is realized. 
If the inequality is violated, the gap opens. 
The same scenario applies when $t_0$, $t_1$ and $t_2$ are replaced by $\tau_0$, $\tau_1$ and $\tau_2$, respectively.

In an H--PTM, arbitrary positive integer ratios of $\tau_{0}, \tau_{1}$ and $\tau_{2}$ can be realized by adjusting the numbers of connections ${N_{\mr{D}0}}, {N_{\mr{D}1}}$ and ${N_{\mr{D}2}}$. 
Therefore, the H--PTM approach can effectively realize various effective honeycomb models with positive integer hopping parameters, 
allowing control of the band gap and the Fermi velocity.

\section{\label{Results}Results}
To validate the proposed low-energy H--PTM-based design method, we compared the bands calculated from the original TB Hamiltonian given by Eq.~(\ref{Ham_H--PTM}) with those of the effective model given by Eq.~(\ref{Heff}) for several cases.

\subsection{\label{R_1}Isotropic case}
\begin{figure}
\includegraphics[width=0.75\linewidth]{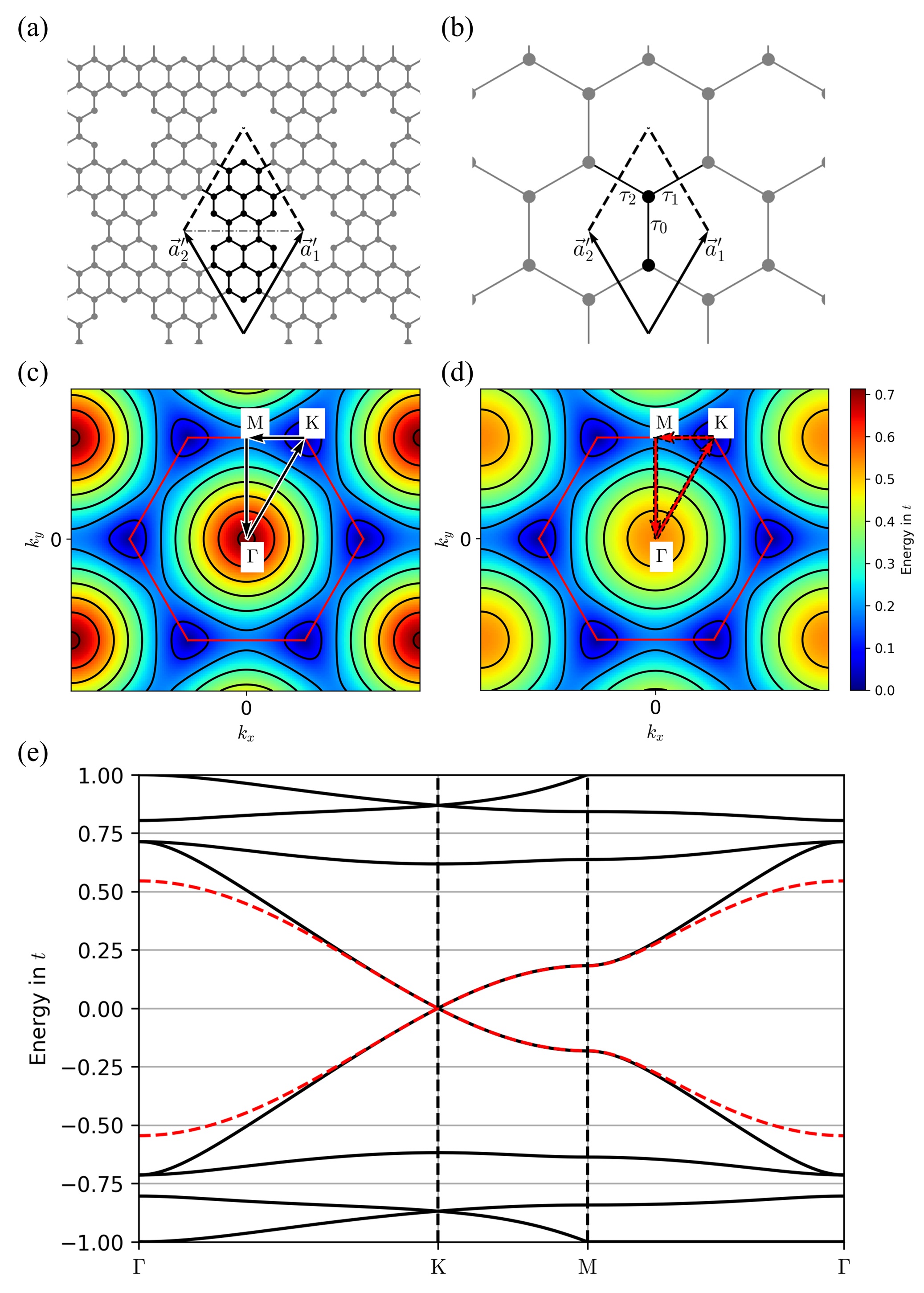}
\caption{\label{Fig4} Results of an isotropic case: (a) The original H--PTM; (b) effective model with $\tau_{0}:\tau_{1}:\tau_{2} =1:1:1$; (c) eigenenergies $|E_{\pm 1}(\vec{k})|$ of the original H--PTM, showing the first Brillouin zone (red hexagon);
(d) eigenenergies $|E_{\mr{eff}}(\vec{k})|$ of the effective model; (e) energy bands along $\Gamma \rightarrow K \rightarrow M \rightarrow \Gamma$. The solid black and dotted red lines show the bands of the original H--PTM and the effective model, respectively.}
\end{figure}

First, let us consider an isotropic case. 
Fig.~\ref{Fig4} (a) shows the most primitive structure of an H--PTM, where each PTM is composed of a single PU.

In this case, $N_{\mr{D}0}:N_{\mr{D}1}:N_{\mr{D}2}=1:1:1$. 
The unit cell forms a regular honeycomb lattice; 
$|\vec{a}'_{1}| = |\vec{a}'_{2}|$ 
and the angle between $\vec{a}'_{1}$ and $\vec{a}'_{2}$ 
is 60$\tcdegree$.

The effective hopping parameters in the effective Hamiltonian take 
$\tau = 2 |{{\varphi}_{\alpha}}||{{\varphi}_{\beta}}| t = 2/11 t \sim 0.1818t$ with $\tau_{0}=\tau_{1}=\tau_{2}=\tau$, 
and the form of the effective model is the same as that of isotropic graphene (Fig.~\ref{Fig4} (b)).
The eigenenergies $E_{\pm 1}(\vec{k})$ of the original H--PTM and the $E_{\mr{eff}}(\vec{k})$ of the effective model shows similar trends throughout the $k$-space (Fig.~\ref{Fig4} (c) and (d)).

Since the parameters $\tau_{0}, \tau_{1}$, and $\tau_{2}$ satisfy the triangular inequality, 
the gap is expected to  close at the energy minimum point, forming an isotropic Dirac cone.
As shown in Fig.~\ref{Fig4} (e), the energy bands of the original model and the effective model exhibit good agreement around the energy minimum point, and a Dirac cone is formed with no gap.

\begin{figure}
\includegraphics[width=0.75\linewidth]{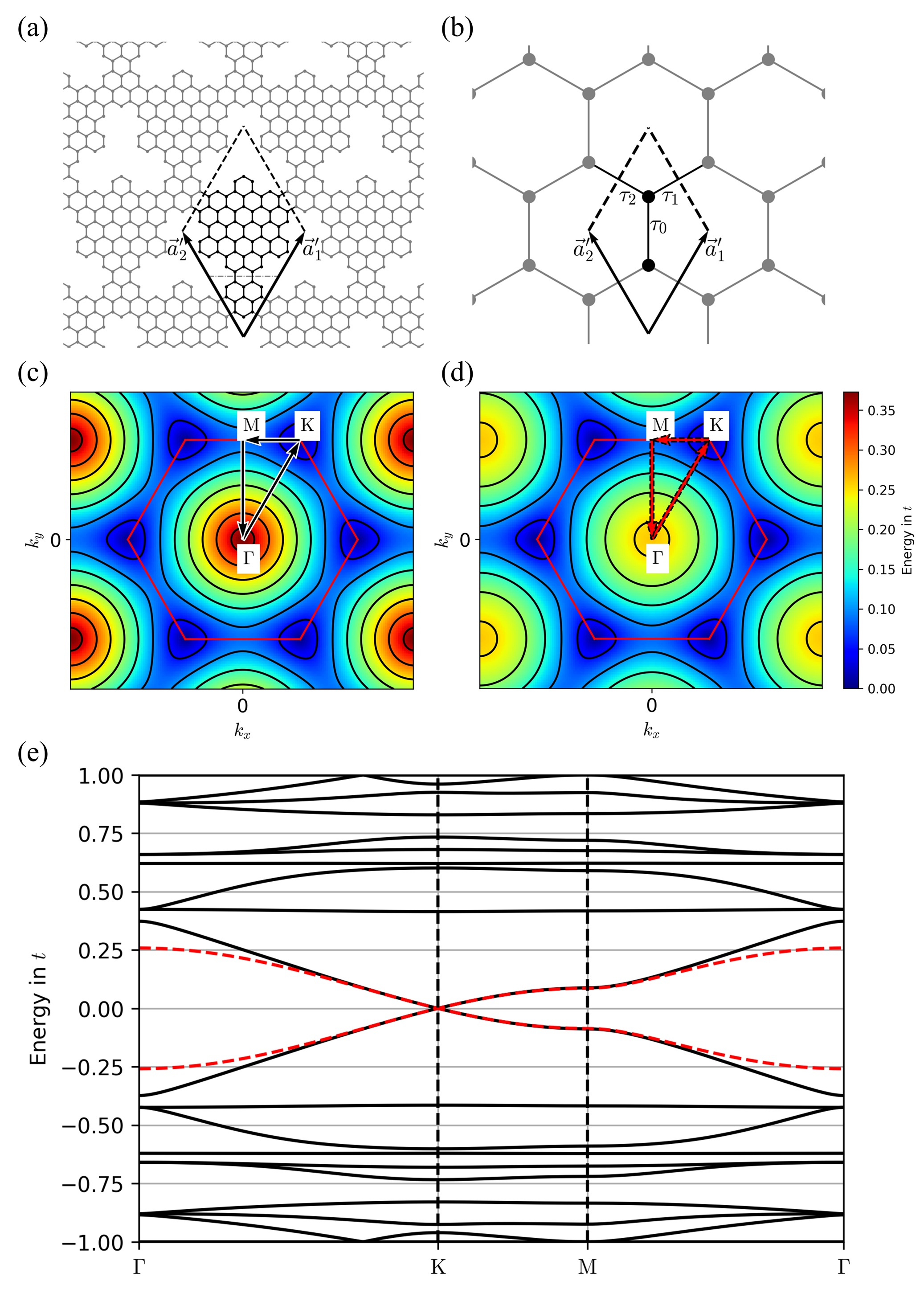}
\caption{\label{Fig5} Another isotropic case with $\tau_{0}:\tau_{1}:\tau_{2} =1:1:1$. (a)-(e) represents the same contents as shown in Fig.~\ref{Fig4}, respectively.}
\end{figure}

We note here that the shapes of the $\alpha$-PTM and $\beta$-PTM can differ. 
Figure.~\ref{Fig5} (a) illustrates another case with $N_{\mr{D}0}:N_{\mr{D}1}:N_{\mr{D}2}=1:1:1$. 
In this case, the $\alpha$-PTM remains a PU while the $\beta$-PTM enlarges. 
However, the hopping parameters $\tau_{0}=\tau_{1}=\tau_{2}=\tau = 6/\sqrt{4843}t \sim 0.0862t$ in the effective model (Fig.~\ref{Fig5} (b)) are isotropic.
The eigenenergies $E_{\pm 1}(\vec{k})$ and $E_{\mr{eff}}(\vec{k})$ of the original H--PTM and effective models are similar throughout the $k$-space (Fig.~\ref{Fig5} (c) and (d)) 
and a Dirac cone is formed as expected (Fig.~\ref{Fig5} (e)).

\subsection{\label{R_2}Critical parameter case}
\begin{figure}
\includegraphics[width=0.75\linewidth]{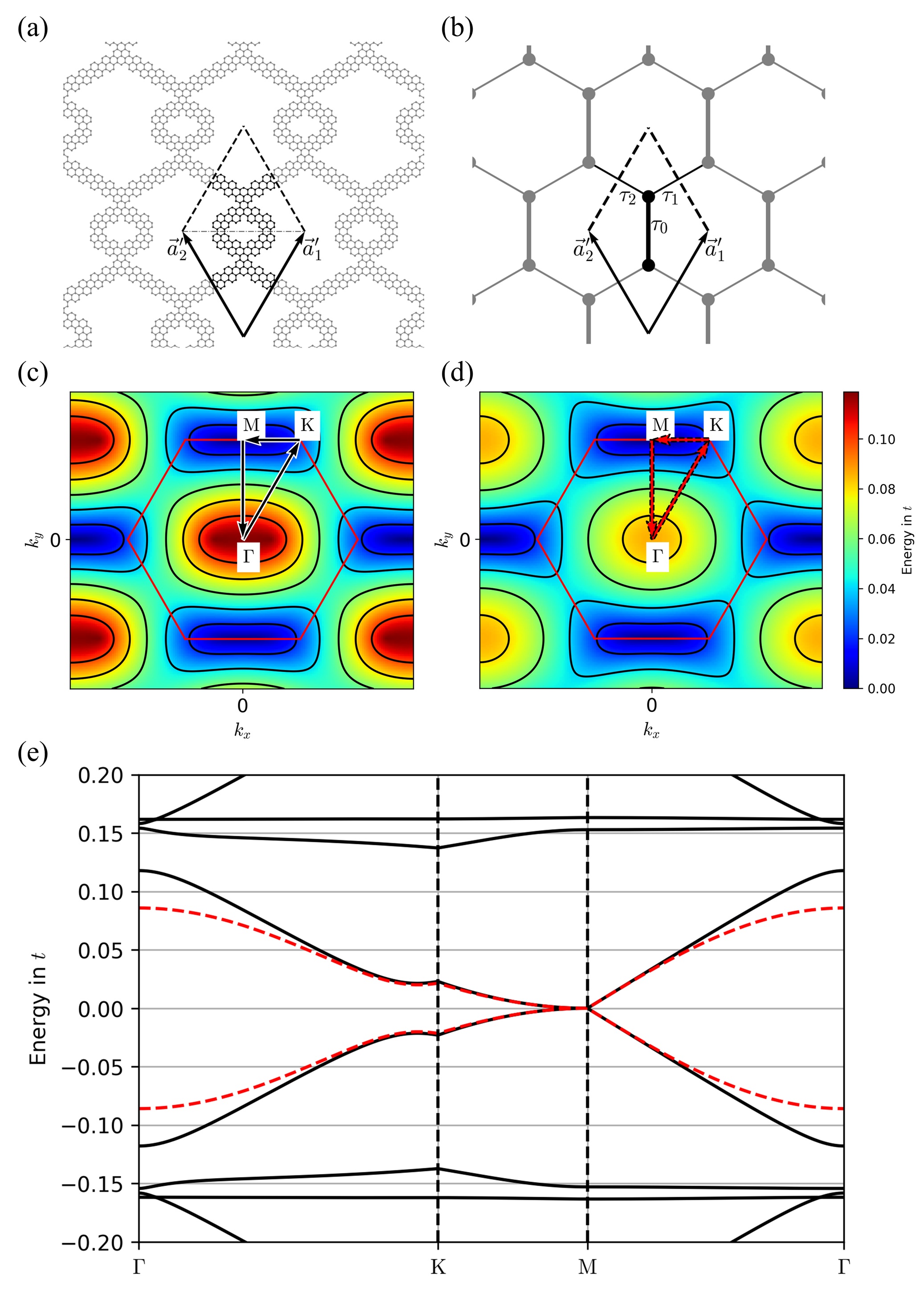}
\caption{\label{Fig6} A critical-parameter case with $\tau_{0}:\tau_{1}:\tau_{2} =2:1:1$.}
\end{figure}

Panels (a) and (b) of Fig.~\ref{Fig6} present the structure of an H--PTM model with $N_{\mr{D}0}:N_{\mr{D}1}:N_{\mr{D}2}=2:1:1$ 
and the corresponding effective model, respectively.

In this case, the parameters $\tau_{0}=2\tau, \tau_{1}=\tau$, and $\tau_{2}=\tau$, where $\tau = 2 |{{\varphi}_{\alpha}}||{{\varphi}_{\beta}}| t = 9/419 t \sim 0.0215t$ in the effective model, are the critical values, 
i.e., $\tau_{0} = \tau_{1} + \tau_{2}$.
The energy minimum in the effective model is expected to move to the confluent point (M-point in this case), where 
the energy gap is just barely closed.

The original H--PTM and effective models yield similar results (Fig.~\ref{Fig6} (c) and (d)) and the band gap in both models is closed at the same confluent point (Fig.~\ref{Fig6} (e)).
In addition, the Fermi velocity is seen to be strongly anisotropic.

\subsection{\label{R_3}Gap-opening case}

\begin{figure}
\includegraphics[width=0.75\linewidth]{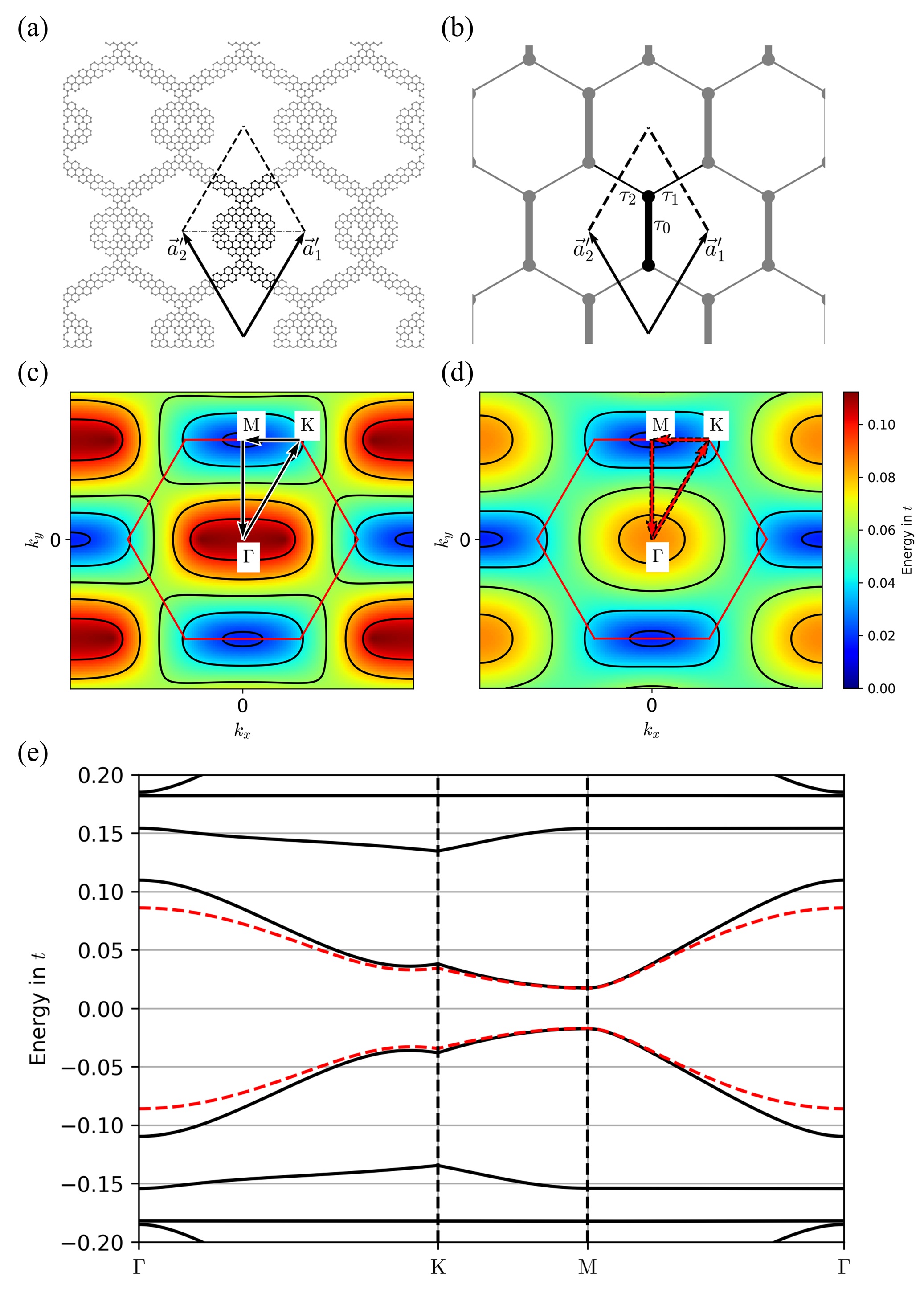}
\caption{\label{Fig7} An anisotropic case with $\tau_{0}:\tau_{1}:\tau_{2} =3:1:1$.}
\end{figure}

In Fig.~\ref{Fig7} (a) and (b), an H--PTM with $N_{\mr{D}0}:N_{\mr{D}1}:N_{\mr{D}2}=3:1:1$ 
and the effective models are shown, respectively.
In this case, since the parameters $\tau_{0}=3\tau, \tau_{1}=\tau$ and $\tau_{2}=\tau$ in the effective model do not satisfy the triangle inequality, 
i.e., $\tau_{0} > \tau_{1} + \tau_{2}$  
a gap is predicted to open at the energy minimum point from the effective model.
In this case, 
since the modes on the A-sites in the $\alpha$-PTM region and on the B-sites in the $\beta$-PTM region 
are modulated from the $\sqrt{3} \times \sqrt{3}$ shape because the eigen energy is not $E=0$. 
So, the value of $\tau = 0.0172t$ in the effective model is numerically determined 
by averaging the values of $|{{\varphi}_{\alpha}}||{{\varphi}_{\beta}}|$ across all DZCs.

We can see the similarity of the original and effective models in Fig.~\ref{Fig7} (c) and (d), 
and the opening of the band gap of both systems in Fig.~\ref{Fig7} (e).

These results demonstrate that the effective model correctly predicts the band openings and closings at different $\tau$ ratios; moreover, 
the effective honeycomb model with integer hopping parameters is precisely realized by the lower-energy part of an H--PTM.

\section{\label{Discussions}Discussion}

\subsection{\label{R_7}Fermi-velocity controllability}

\begin{figure}
\includegraphics[width=1.0\linewidth]{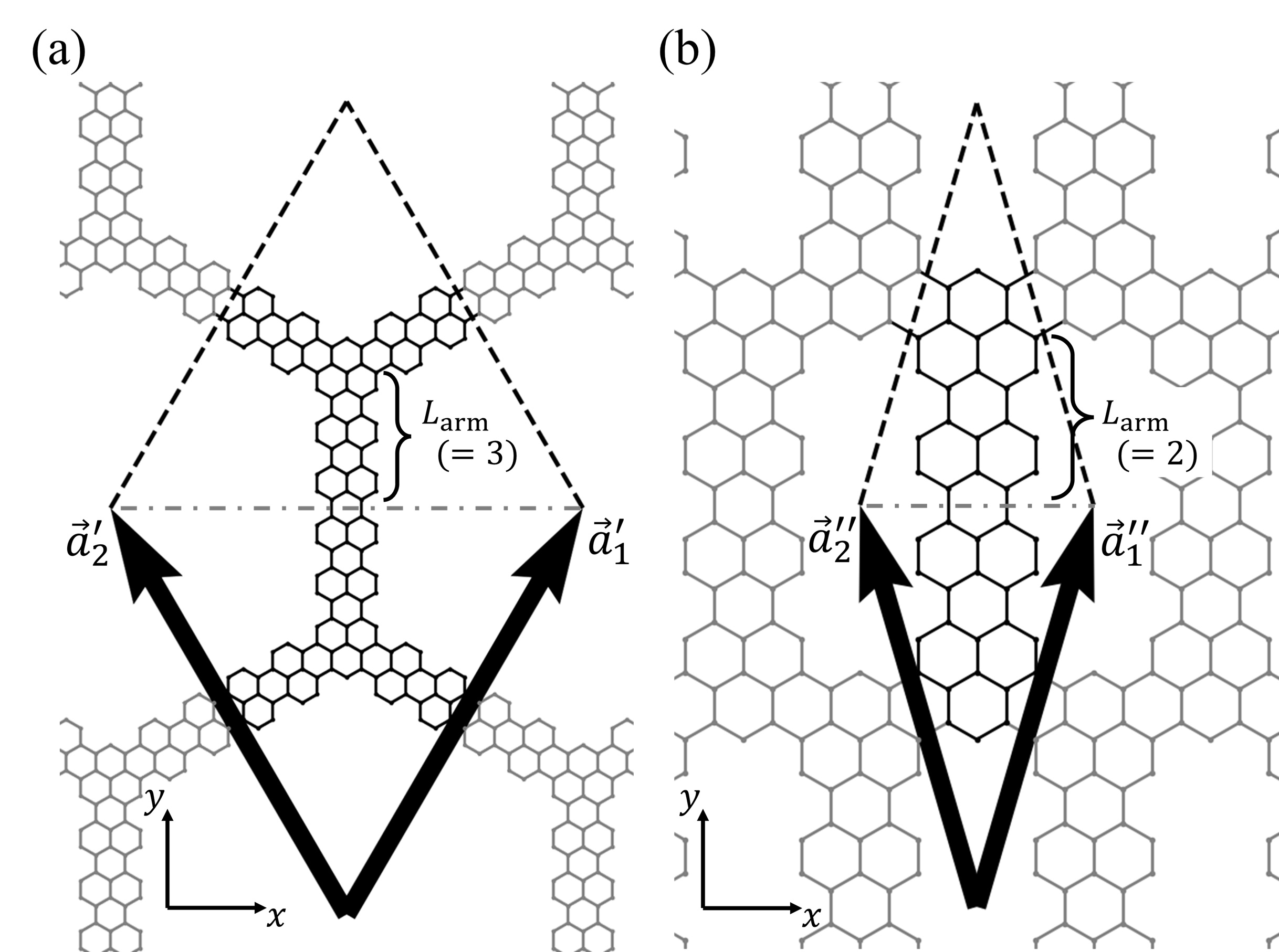}
\caption{\label{Fig8}  H--PTMs composed of a single row of PUs: (a) an isotropic case with $L_{\mr{arm}} = 3$ 
and (b) a uniaxial case with $L_{\mr{arm}} = 2$.}
\end{figure}

This section analyzes the Fermi velocity of the most typical H--PTM example, an H--PTM composed of a single row of PUs.
An isotropic case of this H--PTM is shown in Fig.~\ref{Fig8} (a).
The arms extending in each direction of a PTM are structurally identical to 
the narrowest metallic armchair graphene nanoribbon\cite{PhysRevB.54.17954, PhysRevB.59.8271, enoki2019physics} with a width of $N=5$.
An $N=5$ armchair nanographene has been already synthesized\cite{Kimouche2015-tq}, 
suggesting the feasibility of the present H--PTM structure by bottom-up synthesis techniques\cite{OlympiceneGN, BottomUpSyns}.

Let $L_{\mr{arm}}$ denote the length of each arm.
Each PTM contains $N_\mr{PU} = 3 L_\mr{arm}-2$ PUs.
Under the normalization condition of the eigenvectors, 
$\tau$ can be analytically obtained as follows:
\begin{equation}
  \label{EqDis2}
  \tau = \frac{2}{8 N_\mr{PU}+3}t 
  = \frac{2}{24 L_\mr{arm}-13}t.
\end{equation}
As the lattice vector length is $a' = (6L_{\mr{arm}}-2)a$,
the Fermi velocity $\nu'$ is expressed as 
\begin{equation}
  \label{EqDis3}
  \nu' = \frac{\sqrt{3}}{2} \frac{a' \tau}{\hbar} 
  =\frac{\sqrt{3}}{2} \frac{2(6L_{\mr{arm}}-2)}{24 L_\mr{arm}-13}\frac{a t}{\hbar}.
\end{equation}
Therefore, the Fermi velocity of the H--PTM with respect to the Fermi velocity of graphene is determined as
\begin{equation}
  \label{EqDis4}
  \frac{\nu'}{\nu} = \frac{a'\tau}{at} =\frac{12L_{\mr{arm}}-4}{24 L_\mr{arm}-13}.
\end{equation}
 In the limit $L_\mr{arm} \rightarrow \infty$, we have
\begin{equation}
  \label{EqDis5}
 \frac{\nu'}{\nu} \sim \frac{1}{2}+\frac{5}{48}\frac{1}{L_{\mr{arm}}} \rightarrow \frac{1}{2}. 
\end{equation}

As confirmed in these analyses, an isotropic H--PTM comprising a single row of PUs 
exhibits a maximum Fermi velocity ratio of $\nu'/\nu = 8/11$ at $L_{\mr{arm}}=1$ 
and converges to $\nu'/\nu = 1/2 $ with an order of $1/L_{\mr{arm}}$ as $L_{\mr{arm}}$ increases.

Figure~\ref{Fig8} (b) shows a uniaxial case, which deviates from a regular hexagonal lattice. 
When the arm extends only in the $y$ direction, 
$N_\mr{PU} = L_\mr{arm}$ and 
\begin{equation}
  \label{EqDis6}
  \tau = \frac{2}{8 L_\mr{arm}+3}t 
\end{equation}
is obtained.
Because the lattice vector is uniaxially stretched in the $y$ direction as  
$a''_{1x} = 2a = 4a_x$ and $a''_{1y} = 2\sqrt{3}L_{\mr{arm}}a = 4 L_{\mr{arm}} a_{y}$,
the Fermi velocity ratios in the $x$- and $y$-directions are given by
\begin{equation}
  \label{EqDis7}
  \frac{\nu''_{x}}{\nu} = \frac{a''_{x}\tau}{a_{x}t}=\frac{8}{8 L_\mr{arm}+3} 
\end{equation}
and 
\begin{equation}
  \label{EqDis8}
  \frac{\nu''_{y}}{\nu} 
  = \frac{a''_{y}\tau}{a_{y}t}
  = \frac{8L_{\mr{arm}}}{8 L_\mr{arm}+3}, 
\end{equation}
respectively.
In the limit $L_\mr{arm} \rightarrow \infty$, we have
\begin{equation}
  \label{EqDis9}
 \frac{\nu''_{x}}{\nu} \sim \frac{1}{L_{\mr{arm}}} \rightarrow 0
\end{equation}
and
\begin{equation}
  \label{EqDis10}
 \frac{\nu''_{y}}{\nu} \sim 1-\frac{3}{8} \frac{1}{L_{\mr{arm}}} \rightarrow 1. 
\end{equation}

Therefore, in a uniaxial H—PTM composed of a single row of PUs extending in the $y$ direction alone, 
the Fermi velocity converges to ${\nu''_{x}}/{\nu} = 0 $ and ${\nu''_{y}}/{\nu} = 1$ 
with an order of $1/L_{\mr{arm}}$ as $L_{\mr{arm}}$ increases. 
Note that the Fermi velocity ${\nu''_{y}} = {\nu}$ is the Fermi velocity of infinitely long $N=5$ armchair nanographene\cite{Wakabayashi2010}.

\begin{figure}
\includegraphics[width=1.0\linewidth]{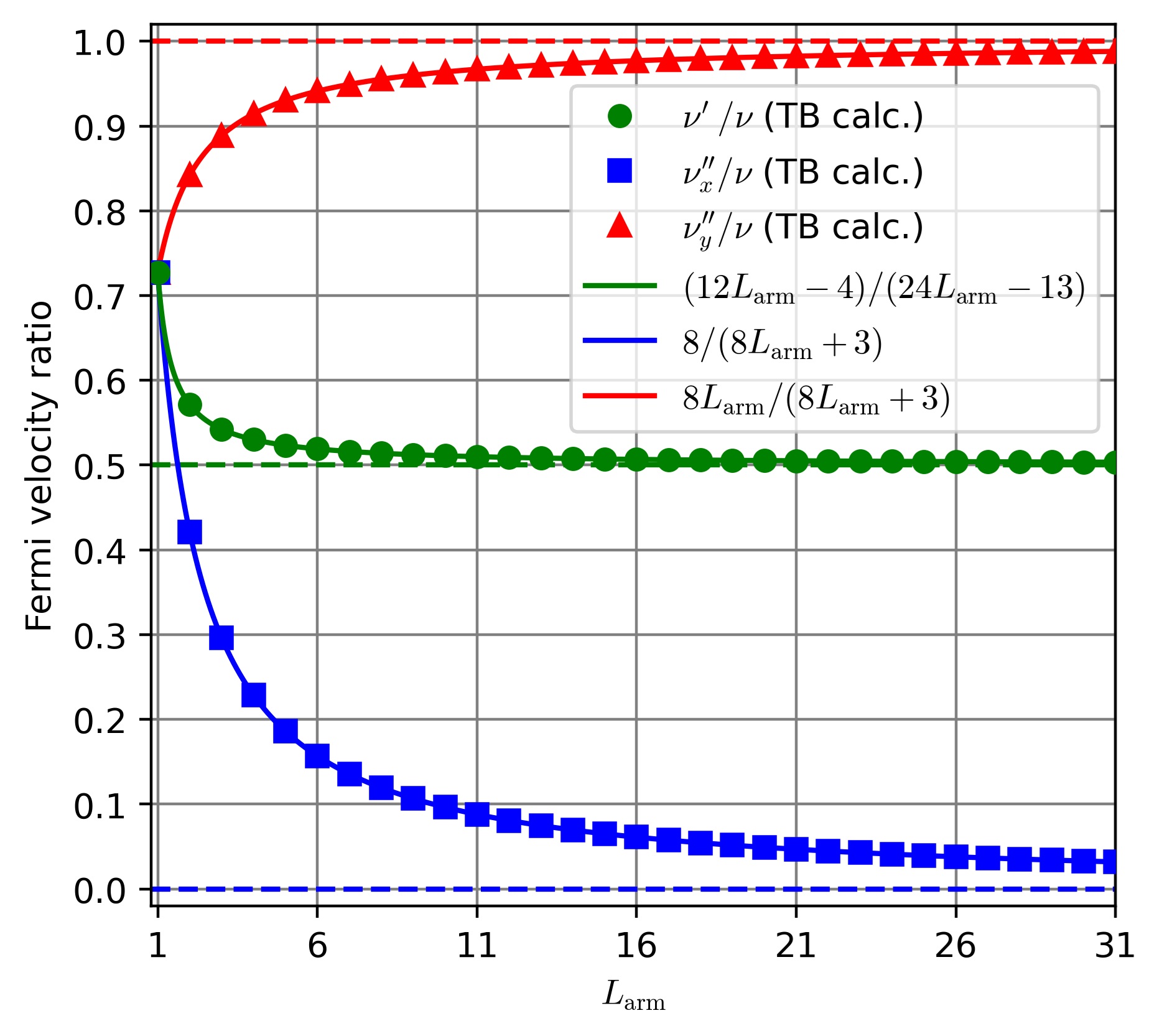}
\caption{\label{Fig9} Arm-length $L_{\mr{arm}}$ dependences of the Fermi velocity ratios ${\nu'}/{\nu}$, ${\nu''_{x}}/{\nu}$ and ${\nu''_{y}}/{\nu}$ relative to the Fermi velocity of graphene. The dots are the values obtained from TB calculations.}
\end{figure}

Figure~\ref{Fig9} shows the $L_{\mr{arm}}$ dependence of the Fermi velocity ratios ${\nu'}/{\nu}$, ${\nu''_{x}}/{\nu}$ and ${\nu''_{y}}/{\nu}$ calculated using Eqs.~(\ref{EqDis4}), (\ref{EqDis7}), and (\ref{EqDis8}). 
The Fermi velocity was controlled over the range 0 to 1$\nu$, and a properly designed H--PTM can realize 
a variety of systems with a Fermi velocity of order comparable to that of graphene.

\subsection{\label{R_6}Vacancies and localized zero modes}
Experiments have shown that hydrogenated carbon defects, 
such as $\mr{V_{111}}$ at single sites can induce localized zero modes.\cite{PhysRevLett.93.187202, PhysRevB.89.155405, doi:10.7566/JPSJ.85.084703}
A similar phenomenon arises from surface hydrogen ad-atoms.\cite{GonzalezHerrero2016}
We can consider that vacancies can be introduced in each PTM without violating the PU tessellation rule, 
i.e., to the extent that each PTM can be represented by a tiling of $\alpha$-PUs or $\beta$-PUs.

\begin{figure}
\includegraphics[width=1.0\linewidth]{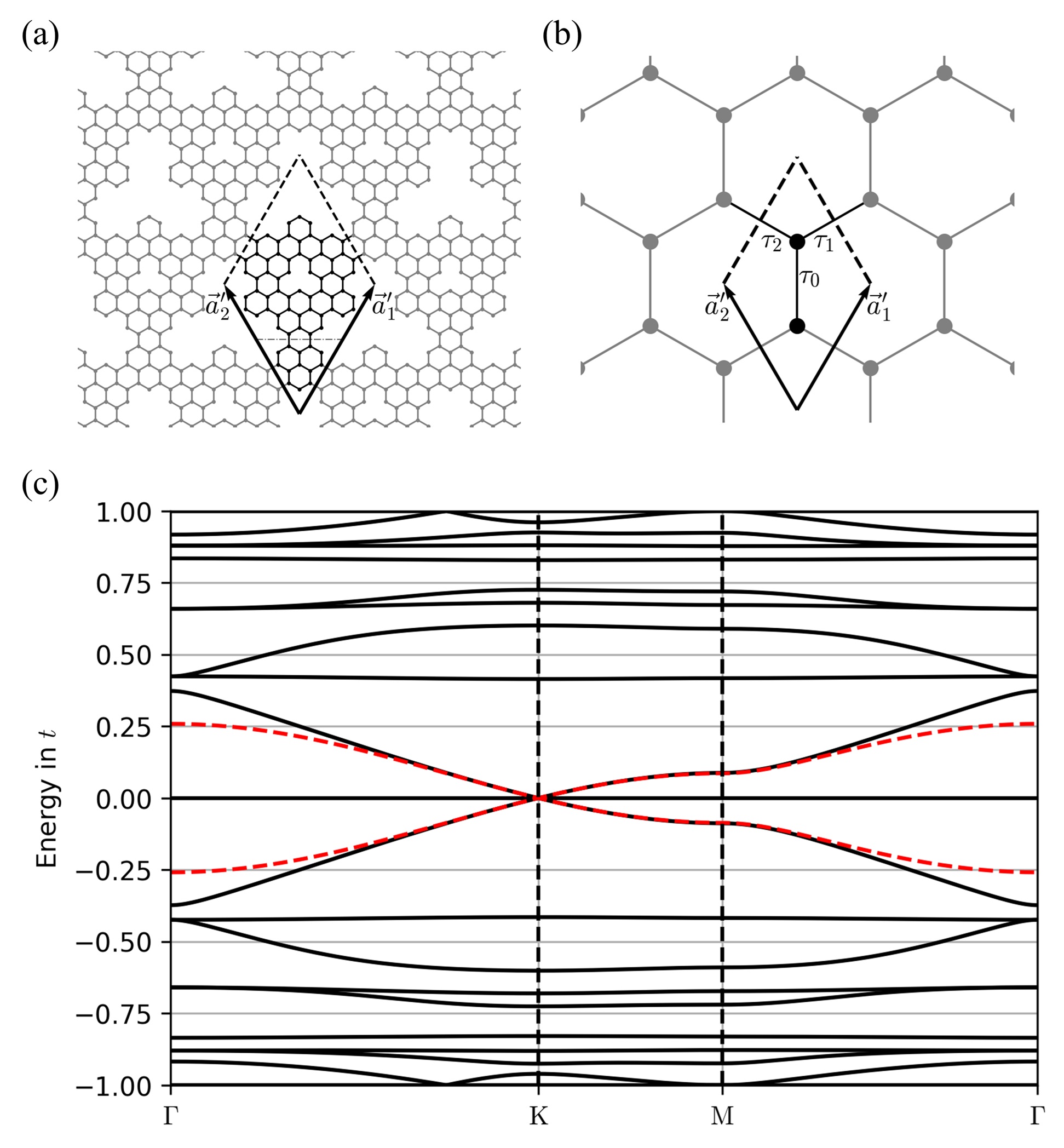}
\caption{\label{Fig10} Introduction of a vacancy in a $\beta$-PTM ($N=0$ and $M=1$ case): 
(a) H--PTM model with a vacancy in the $\beta-$PTM, (b) effective model of the Dirac modes $E_{\mr{eff}}(\vec{k})$, and (c) energy bands of the original model (solid black lines) and the effective model (red-dashed lines). 
A zero mode appears $E = 0$ without changing the Dirac cone seen in Fig.~\ref{Fig5} (e).} 
\end{figure}

Figure~\ref{Fig10} (a) shows the same H--PTM as in Fig.~\ref{Fig5} but with 
a single vacancy at the center of the $\beta-$PTM.  
The effective model of the Dirac mode (Fig.~\ref{Fig10} (b)) remains unchanged from that of Fig.~\ref{Fig5} (b).
The energy band of this H--PTM with a vacancy (Fig.~\ref{Fig10} (c)) displays a Dirac cone 
and an additional zero localized mode at $E=0$ originating from the coexisting vacancy.
In fact, it can be shown that even when vacancies are introduced at the center of PUs, 
the shape of the vacancy edges matches the shape of the $\sqrt{3} \times \sqrt{3}$ mode as far as each PTM is constructed by a tiling of $\alpha$-PUs or $\beta$-PUs.\cite{MORISHITA2021PLA}
Therefore, the $\sqrt{3} \times \sqrt{3}$ mode remains an eigenstate after the vacancies are introduced. 
A variety of shapes and multiple numbers of vacancies can be introduced\cite{MORISHITA2021PLA, Morishita_APEX_2021, KomatsuJPSJ2025} 
without disturbing any results on the effective $\sqrt{3} \times \sqrt{3}$ zero modes discussed in previous sections. 
This is a revised PU tessellation rule certifying coexisting Dirac modes with the localized zero modes.

In general, when $N$ and $M$ vacancies are introduced into the $\alpha$-PTM and $\beta$-PTM, respectively, 
the resulting vacancy-induced zero modes are localized within their respective PTM region, \cite{doi:10.7566/JPSJ.88.124707, MORISHITA2021PLA} 
and a total of $N+M$ zero modes appear at $E=0$.
Figure~\ref{Fig10} illustrates a case of $N=0$ and $M=1$.
In a recent work, we have also studied specific cases of $N=M=3$ and confirmed that a total of six vacancy-induced zero modes appear.\cite{KomatsuJPSJ2025}

The highly localized zero modes embedded in the system are expected to induce a strong correlation effect. 
The number of electrons is enough to fill the localized zero modes just at the half-filling. 
Then, the zero mode orbitals in each PTM are expected to be singly-occupied. 
The strong correlation effect should appear stronger in the localized zero mode than in the itinerant Dirac band. 
For example, even when a weak correlation effect is considered in the case of Fig.~\ref{Fig10},  
the flat band associated with the localized zero modes exhibits spin characteristics with $S = 1/2$ at each zero mode.

An effective approach to predict magnetic phases using the Hubbard model having flat bands\cite{PhysRevLett.72.144} 
is applicable by treating the whole $\pi$ system, too. 
Depending on the values of $N$ and $M$, ferrimagnetic systems may appear. 
When the model includes not only localized modes but also Dirac bands that can be modulated, 
systems of localized spins interacting with itinerant electrons can be realized.

Consequently, H--PTM-based design has the potential to realize not only high-spin states,
but also more diverse electron systems in which interacting localized electron spin coexist and interact with various types of Dirac electron systems with or without a gap.

\section{\label{Conclusion}Conclusion}
We have proposed a class of nanographene effective atom structures called honeycomb phenalenyl-tessellation molecules (H-PTMs) 
formed through the systematic tessellation of phenalenyl units (PUs) and theoretically analyzed their low-energy band structures.
Our findings show that a very simple two-site effective model can capture the low-energy band behavior of an H-PTM, 
and the effective hopping parameters in the model can be designed with an arbitrary positive integer hopping ratios.
The opening and closing of band gaps can also be controlled by designing the effective hopping parameters under the triangular inequality criterion. 
The isotropic and anisotropic Fermi velocities can also be designed over a wide range, from zero to a value comparable to that of pristine graphene when the gap closes.
Furthermore, vacancies can introduce zero modes into an H--PTM, enabling the design of
coexisting systems of modulated Dirac bands and localized zero modes.
This study demonstrates that H--PTMs can significantly extend the design freedom of electronic states from those of pristine graphene, 
enabling the design of graphene-based electronic and quantum devices with precisely controlled conduction electrons and localized spins. 
Finally, the unique features of H--PTMs and related molecular architectures can be exploited for the design of highly specialized graphene-based devices.

\begin{acknowledgments}
The authors thank T. Yamaguchi and Y. Oishi for valuable discussions. 
This work was supported by JSPS KAKENHI Grant Numbers 
JP22K04864, JP23H03817 and JP24K17014.
\end{acknowledgments}

\section*{Data Availability}
The data that support the findings of this article are openly available.\cite{materialscloud}

\appendix*
\section{}
\renewcommand{\thefigure}{A.\arabic{figure}}
\setcounter{figure}{0}
The analytical solution of the wavefunction on the A-sites at the K-point for the system shown in Fig.~\ref{Fig4} is shown in Fig.~\ref{Fig11(A1)} (a). 
$\omega = e^{\frac{2}{3} \pi i}$, $\omega^2 = e^{\frac{4}{3} \pi i}$ are the cube roots of unity. 
From symmetry, the wavefunction on the B-sites has the same structure.
$|{{\varphi}_{\alpha}}|$ and $|{{\varphi}_{\beta}}|$ are determined by the normalized analytical solution; 
$|{{\varphi}_{\alpha}}| = |{{\varphi}_{\beta}}| = 1/\sqrt{11}$.
Therefore, $\tau = 2|{{\varphi}_{\alpha}}||{{\varphi}_{\beta}}|t=2/11t$.

Similarly, the analytical solution on the A-sites at the M-point of the system in Fig.~\ref{Fig6} is also shown in Fig.~\ref{Fig11(A1)} (b).
In this case, $|{{\varphi}_{\alpha}}| = |{{\varphi}_{\beta}}| = 3/\sqrt{838}$ and $\tau = 2|{{\varphi}_{\alpha}}||{{\varphi}_{\beta}}|t=9/419t$.

The analytical solutions on the A-sites and the B-sites of the system in Fig.~\ref{Fig5} are shown in Fig.~\ref{Fig11(A1)} (c) and (d), respectively.
In this case, $|{{\varphi}_{\alpha}}| = 3/\sqrt{167}$ and $|{{\varphi}_{\beta}}| = 1/\sqrt{29}$.
Therefore, $\tau = 2|{{\varphi}_{\alpha}}||{{\varphi}_{\beta}}|t=6/\sqrt{4863}t$.

For systems with arms consisting of a single row of PUs, 
such as shown in Fig.~\ref{Fig3} and Fig.~\ref{Fig8}, 
the analytical solution can be easily obtained by simply considering a wavefunction that spreads with uniform amplitude on the arms, 
based on Fig.~\ref{Fig11(A1)} (a).

\begin{figure*}
\includegraphics[width=1.0\linewidth]{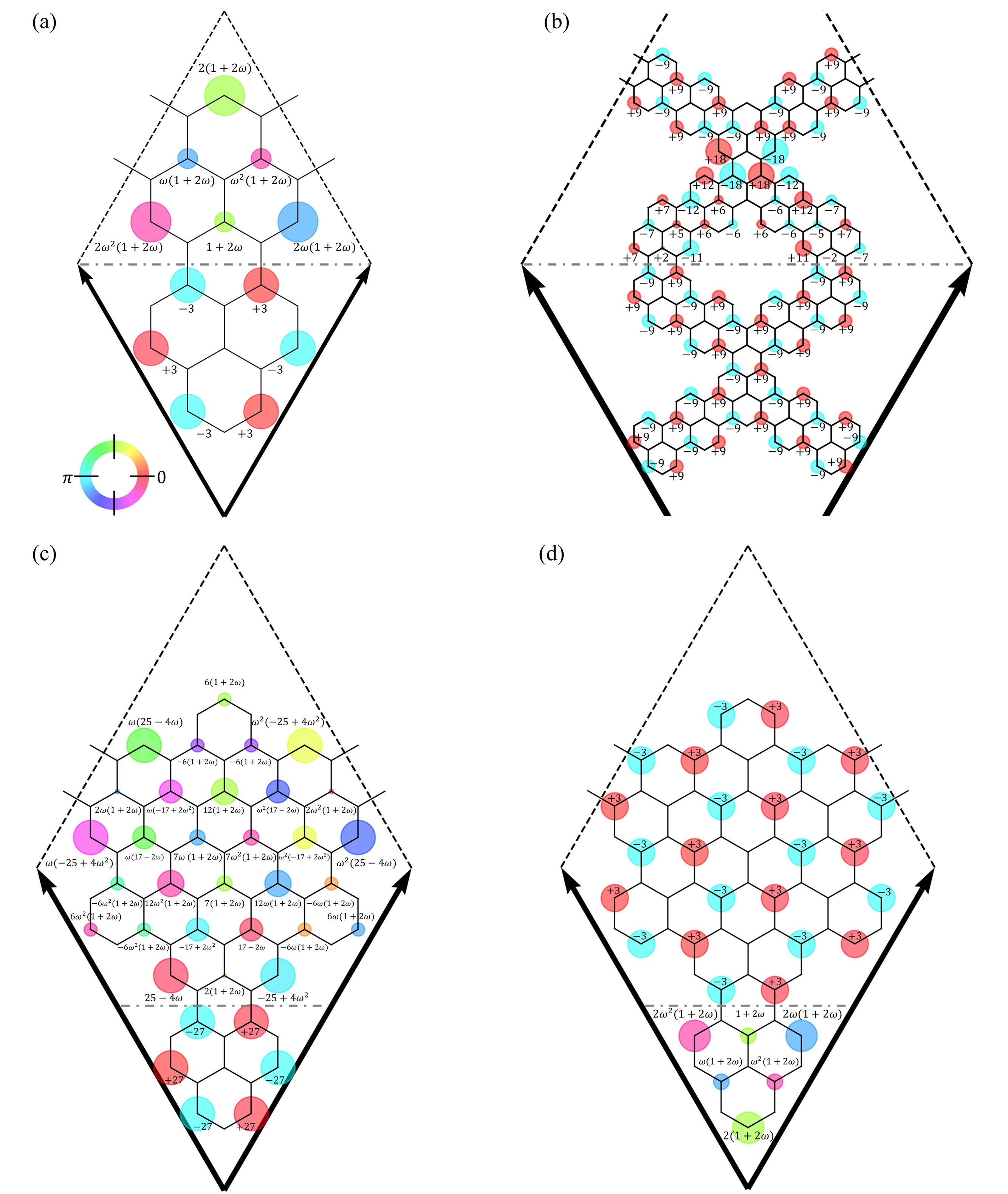}
\caption{\label{Fig11(A1)} The analytical solutions of the wavefunctions for the system shown in (a) Fig.~\ref{Fig4} (A-sites, K-point), 
(b) Fig.~\ref{Fig6} (A-sites, M-point), 
(c) Fig.~\ref{Fig5} (A-sites, K-point) and 
(d) Fig.~\ref{Fig5} (B-sites, K-point).}
\end{figure*}

\bibliography{ModulatedDirac}

\end{document}